\documentstyle[epsfig]{aipproc}

\begin{document}
\title{$\sigma$-Particle in Production Processes}

\author{Kunio Takamatsu$^*$, Muneyuki Ishida$^{\dagger}$,
Shin Ishida$^{\ddagger}$\\ 
Taku Ishida$^{\star}$ and
Tsuneaki Tsuru$^{\star}$}
\address{$^*$Miyazaki U., Gakuen-Kibanadai, Miyazaki 889-21,
Japan\\
$^{\dagger}$Department of Physics, U. of Tokyo, Hongou, Bunkyo, 
Tokyo 113, Japan\\
$^{\ddagger}$Coll. Sci. and Tech. Nihon U., Kanda-Surugadai, 
Chiyoda, Tokyo 101, Japan\\
$^{\star}$KEK, Oho, Tsukuba, Ibaraki 305, Japan}

\maketitle

\begin{abstract}
A $S$-wave $\pi\pi$ resonance below 1 GeV/c$^2$, 
$sigma$ has been observed in analyses 
of the phase shift data of $\pi\pi$ scatterings. 
It is important to observe it in production 
processes. A huge events of $\pi^0\pi^0$ below 1 GeV/c$^2$ is seen 
in the central production of the 
GAMS experiment. Its mass spectrum is well described 
by the variant mass and width 
method (VMW) including $f_0$(980), 
$f_2$(1270) and a $S$-wave resonance which might be 
assigned to be a sigma. 
We report here that the angular distributions 
of the $\pi^0\pi^0$ system are also described with
the $S$-wave resonance interfering with 
$f_0$(980) and the tail of 
$f_2$(1270) by VMW. 
The same method is applied to the  
$\pi^+\pi^-$ data of 
$J/\psi\rightarrow\omega\pi\pi$ decay, and is shown that not 
only $\pi^+\pi^-$ mass spectrum but also its 
$cos\theta_\pi$ distributions are described well.
\end{abstract}

We have reported\cite{rf:1} that 
a $S$-wave $\pi\pi$ resonance is observed 
below 1 GeV/c$^2$ in the reanalysis 
of the phase shift data\cite{rf:3} of 
the $\pi\pi$ scatterings. 
Several analyses on the same data also have reported\cite{rf:4} 
an exsistence of a possible scalar resonance(s) 
of the $\pi\pi$ system. An existence of a sigma 
particle is expected in this region as a chiral partner of
Nambu-Goldstone boson $\pi$. Its mass value 
is predicted to be twice of 
constituent mass of a $n$ quark
in the extended Nambu-Jona Lasinio model\cite{rf:5} as the 
low energy effective theory of QCD. 
It is vitally important to observe the sigma in 
production processes as well as in the scattering process.

\begin{figure}[htb]
  \epsfysize=6. cm
  \centerline{\epsffile{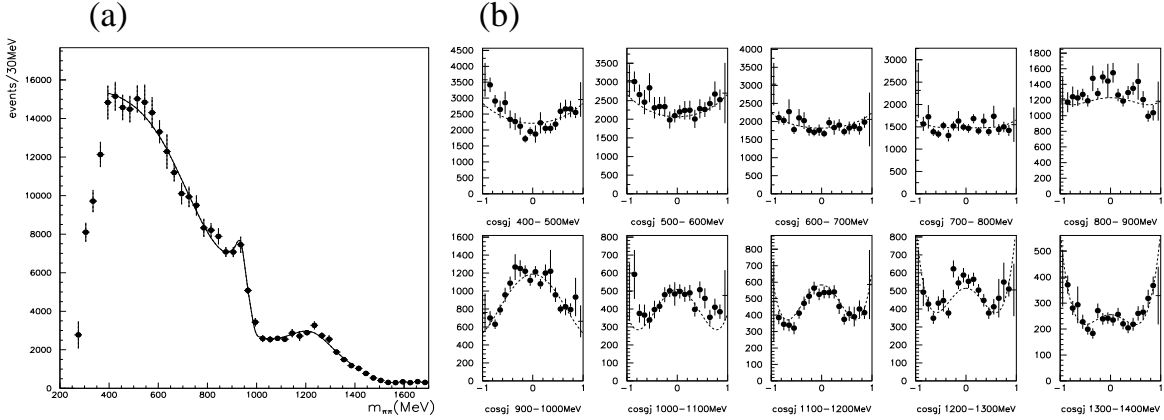}}
    \caption{(a) $\pi^0\pi^0$ mass spectrum (corrected for detection
    efficiency). The solid line represents the fit by
    two Breit-Wigners. 
    (b) $cos\theta_{GJ}$ distribution for various mass interval.
     The result of the fit is shown by dashed lines.}
    \label{fig:1}
\end{figure}
The GAMS group has observed 
huge $\pi^0\pi^0$ events below 1 GeV/c$^2$ in a proton 
proton central collision process at 450 GeV/c.
Details of the experimental condition 
and data are described in reference \cite{rf:6}. 
It shows a resonant peak at 500-600 MeV/c$^2$,
as shown in Fig.\ref{fig:1}(a). It can be described 
by a $S$-wave Breit-Wigner interfering with $f_0$(980).
The variant mass and width method(VMW)\cite{rf:10}
is applied to the mass spectrum.
Here we summarize formulae of the analysis, briefly.
An amplitude, ${\cal M}$, is given by the product 
of production amplitude, ${\cal M}^{prod}(f_j)$, 
propagator $\Delta^j$, and coupling constant $g_{j\pi\pi}$. 
$j$ indicates $j$-th resonance.
$f_0$(980), $f_c$ and $f_2$(1270) are supposed for $j$, 
where $f_c$ means a $\pi\pi$ resonance below 1 GeV/c$^2$. 
Helicity couplings, $\rho_h$ ($h$= -2 to+2, $\sum\rho_h^2$= 1) 
is considered for $f_2$. 
An introduction of relative phases is an important and essential point 
in the VMW method. We define
production amplitude, $\xi$, as 
${\cal M}^{prod}(f_j)$$\Delta^j g_{j\pi\pi}$
$\equiv$$\xi_j e^{i\theta_j}$. 
$\theta_{f_0}$ is taken to be $zero$ by definition. 
The total amplitude is written as follows,
\begin{eqnarray}
{\cal M}&=&
\frac{\xi_{f_0}}{(s-m_{f_0}^2)+i\sqrt{s}\Gamma_{f_0}^{tot}(s)}
+
\frac{\xi_{f_c}e^{i\theta_{f_c}}}{(s-m_{f_c}^2)+i\sqrt{s}\Gamma_{f_0}^{tot}(s)}
\nonumber \\
&+&
\frac{\xi_{f_2}e^{i\theta_{f_2}}}{(s-m_{f_2}^2)+i\sqrt{s}\Gamma_{f_2}^{tot}(s)}
\sum_h \rho_h \varepsilon_{\mu\nu}^{(h)}
\frac{(p_1 -p_2)_\mu}{\sqrt{s}}
\frac{(p_1 -p_2)_\nu}{\sqrt{s}}.
\label{eq:1}
\end{eqnarray}
${\cal M}$ is modified by $(s-m_\pi^2)/s\cdot{\cal M}$ to take
Adler zero into account.
Differential Pomeron-Pomeron 
cross section and total cross section are 
\begin{eqnarray}
d\sigma_{PP}=\frac{|{\bf p_1}|}{16\pi^2\sqrt{s}}
|{\cal M}|^2 d\Omega_1
\label{eq:2}
\end{eqnarray}
(${\bf p_1}$ being momentum in the $2\pi$ rest frame) and
\begin{eqnarray}
\sigma_{PP}&=&\frac{|{\bf p_1}|}{4\pi\sqrt{s}}
\left\{
\frac{\xi(f_0)^2}{(s-m_{f_0}^2)^2+s(\Gamma_{f_0}^{tot}(s))^2}\right.
\nonumber \\
&+&\frac{\xi(f_c)^2}{(s-m_{f_c}^2)^2+s(\Gamma_{f_c}^{tot}(s))^2}
+
\frac{\xi(f_2)^2}{(s-m_{f_2}^2)^2+s(\Gamma_{f_2}^{tot}(s))^2}
\cdot\frac{32}{15}\frac{|{\bf P_1}|_{\pi\pi}^4}{s^2}\nonumber \\
&+&\left.\left[\frac{\xi(f_c)\xi^*(f_0) e^{i\theta_{f_c}}}
{(s-m_{f_c}^2+i\sqrt{s}\Gamma_{f_c}^{tot}(s))
(s-m_{f_0}^2-i\sqrt{s}\Gamma_{f_0}^{tot}(s))}\ +\ c.c.\right]\right\}.
\label{eq:3}
\end{eqnarray}
In the actual analyses for production processes,
appropriate kinematical factor is multiplied to Eq.~(\ref{eq:3}).

The solid line in Fig.~\ref{fig:1}(a) shows the result 
of fit by Eq.\ref{eq:3}). Following parameters for 
mass and width have been obtained for $f_c$;
\footnote{
In ref.\cite{rf:6} $im\Gamma(s)$ was used instead
of $i\sqrt{s}\Gamma(s)$ in Eq.~(\ref{eq:1}), and gives 
a heavier $\sigma$ mass.
}
\begin{eqnarray*}
m_c&=&590\pm10\ {\rm MeV/c}^2,\ \ \ 
\Gamma_{c\pi\pi}=710\pm30\ {\rm MeV/c}^2.\\
\end{eqnarray*}
The $cos\theta_{GJ}$ angular distributions of the $\pi^0\pi^0$ system 
are also fitted with Eqs.~(\ref{eq:1}) 
and (\ref{eq:2}), which are shown 
in fig.~\ref{fig:1}(b) for every 100 MeV/c$^2$ mass interval.
Dotted curves in the figures are results of the fit. 
Parameters for masses and others are common to
those obtained above. The values of 
$\theta_{f_c}$ and $\theta_{f_2}$ are 210 and 235 in degrees, 
respectively. The $\pi^0\pi^0$ data are described excellently by VMW not
only for the mass distribution, but also for angular distributions.
\footnote{It is to be noted that
M.~Pennington claimed
at the Hadron'95\cite{rf:7} that our reported result of
the analysis of the $\pi^0\pi^0$ mass spectrum produced in the $pp$ 
central collision is not acceptable. 
This claim was caused from inappropriate application 
of ``pole-universality'' 
in his (their) analysis of phase shifts of the 
$\pi\pi$ scattering data\cite{rf:8}. 
Detailed discussion on the ``universality" 
has been given at this conference\cite{rf:9}.
}

\begin{figure}[t]
  \epsfysize=7.5 cm
 \centerline{\epsffile{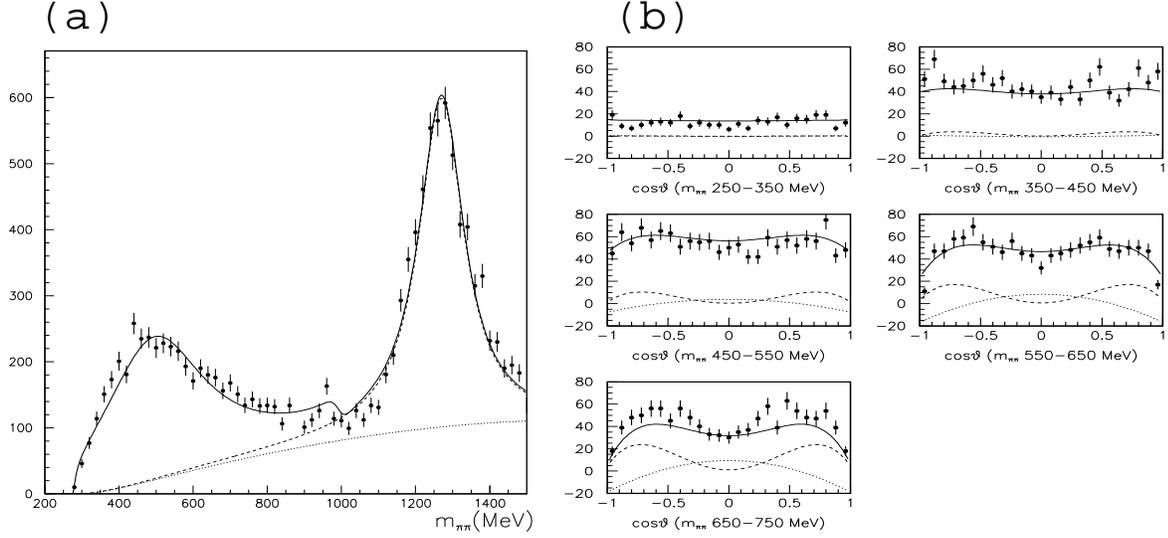}}
 \caption{(a) The fit for effective mass distribution
 by VMW-method shown by a solid line. The dashed line is
 $D$-wave contribution coming from $f_2$($D_{f_2}$) and 
 $b_1\pi$ decay($D_{BG}$).
 Dotted line shows $D_{BG}$ contribution. 
 (b) The fit for $cos\theta_\pi$ distribution
  by VMW-method shown by a solid line. The dashed lines are
  $D$-wave contribution $|D_{f_2}|^2$+$|D_{BG}|^2$, while 
  dotted lines are interference of $f_2$ and $S$-wave state,
  $|S\cdot D_{f_2}|$. Data are taken from ref.~10).}
  \label{fig:2}
\end{figure}
We have performed an anlysis on 
$\pi^+\pi^-$ data of $J/\psi\rightarrow\omega\pi\pi$\cite{rf:11} 
obtained by the DM2 collaboration, which also shows a huge 
events in low mass region.
It is reported that they obtained 
$m_{low~mass}$= 414$\pm$ 20 MeV/c$^2$ 
and $\Gamma_{low~mass}$=
494 $\pm$ 58 MeV/c$^2$ for the low mass $S$-wave by fit with 
two BW's + polynomials, ignoring their mutual interferences. 
The fit could not reproduce 
the $cos\theta_\pi$ distributions
in the mass region between 550 and 750 MeV/c$^2$, 
as is mentioned in their report. 

An estimation of a contribution of $\pi^+\pi^-$ bacground events
coming from $J/\psi\rightarrow b_1(1235)\pi$ and 
$b_1(1235)\rightarrow\pi^+\pi^-\pi^-\pi^0$, which 
shows a $D$-wave like behavior,
is important in the analysis.
We will take this effect $D_{BG}$ as follows;
\begin{eqnarray}
|{\cal M}|^2 &=& |S+D_{f_2}|^2 + |D_{BG}|^2,\nonumber\\
|D_{BG}|^2&=&a_0^2\frac{p_1^4}{s^2}(Y^{(0)})^2 +
2a_1^2\frac{p_1^4}{s^2}(Y^{(1)})^2 +
2a_2^2\frac{p_1^4}{s^2}(Y^{(2)})^2,
\end{eqnarray} 
with parameters $a_i$($i$=0,1,2).
Fig.~\ref{fig:2}(a) shows the experimental data 
of the $\pi^+\pi^-$ mass distribution, where
a solid curve is the result 
of fit of the VMW. 
$m_{low\ mass}$=482$\pm$3 MeV/c$^2$ and
$\Gamma_{low\ mass}$=325$\pm$10 MeV/c$^2$
are obtained. 
The dashed curve in the figure 
indicates the contribution from the $D$-waves of both 
$f_2$(1270) and $b_1\pi$. 
A dotted curve indicates that of $b_1\pi$. 

The $cos\theta_\pi$ distributions 
are also fitted, as shown in 
fig.~\ref{fig:2}(b) in every 100 MeV/c$^2$ mass interval. 
Solid curves in the figures show results. 
Dashed curves in the figure indicate the contribution from the $D$-waves. 
$\theta_{low\ mass}$=214 and 
$\theta_{f_2}$=157 in degrees are obtained for relative 
phases between the low mass and $f_2$, respectively. 
The qualitative features of data are reproduced well by VMW, by
including interferences between the $S$-wave andS tail of $D$-wave 
from $f_2$(1270), whose contributions are shown by dotted lines 
in the figures.


\begin{references}
\bibitem{rf:1}S. Ishida et al., 
   {\it Prog. Theor. Phys.} {\bf 95}, 745 (1996); 
   {\it Prog. Theor. Phys.} {\bf 98}, 1005 (1997).
    T.~Ishida, this conference.
\bibitem{rf:3}B. Hyams et al., 
   {\it Nucl. Phys.} {\bf B64}, 134 (1973). 
   G. Grayer et al., {\it Nucl. Phys.} {\bf B75}, 189 (1974).
\bibitem{rf:4}N.N.~Achasov and G.N.~Shestakov, 
   {\it Phys. Rev.} {\bf D49}, 5779 (1994). 
   R.~Kami\'nski, L.~Le\'sniak and J.-M.~Maillet, 
   {\it Phys. Rev.} {\bf D50}, 3145 (1994). 
   N.A..~T\"ornqvist and M.~Roos, 
   {\it Phys. Rev. Lett.} {\bf 76}, 1975 (1996). 
   M.~Harada, F.~Sannino and J.~Schechter,
    {\it Phys. Rev. D} {\bf 54}, 1991 (1996).  
\bibitem{rf:5}R.~Delbourgo and M.D.~Scadron, 
    {\it Phys. Rev. Lett.} {\bf 48}, 379 (1982). 
   T.~Hakioglu and M.D.~Scadron, 
    {\it Phys. Rev. D} {\bf 42}, 94 (1990). 
   T.~Hatsuda and T.~Kunihiro, 
    {\it Prog. Theor. Phys.} {\bf 74}, 765 (1985); 
    {\it Phys. Rep.} {\bf 247}, 221 (1994).
\bibitem{rf:6}D.~Alde et al., 
    {\it Phys. Lett.} {\bf B397}, 350 (1997).
\bibitem{rf:10}S.~Ishida et al., 
  {\it Prog. Theor. Phys.} {\bf 88}, 89 (1992).
\bibitem{rf:7}M.R.~Pennington, 
    {\it Proc. 6th Int. Conf. Hadron Spectroscopy, Hadron '95}, 
      Manchester, 1995, ed. by M.C. Birse et al. (World Scientific) p.3.
\bibitem{rf:8}K.L.~Au, D.~Morgan and M.R.~Pennington, 
    {\it Phys. Rev.} {\bf D35}, 1633 (1987). 
   D.~Morgan and M.R.~Pennington, 
    {\it Phys. Rev.} {\bf D48}, 1185 (1993).
  \bibitem{rf:9}S.~Ishida, this conference. 
  M.Y.~Ishida, this conference.
\bibitem{rf:11}J.E.~Augustin et al., 
  {\it Nucl. Phys.} {\bf B320}, 1 (1989).
\end{references}
\end{document}